\documentclass[usenatbib, referee]{mn2e}
\usepackage{graphicx}
\usepackage{epsfig}
\usepackage{color}

\title[UCXBs with He star companions]
{Ultra-compact X-ray binaries with He star companions}
\author[Wang et al.]
{Bo Wang,$^{\rm 1}$\thanks{E-mail:wangbo@ynao.ac.cn} 
Wen-Cong Chen,$^{\rm 2,3}$\thanks{E-mail:chenwc@pku.edu.cn}
Dong-Dong Liu,$^{\rm 1}$ 
Hai-Liang Chen,$^{\rm 1}$  
\newauthor
Cheng-Yuan Wu,$^{\rm 4}$ 
Wen-Shi Tang,$^{\rm 5}$
Yun-Lang Guo$^{\rm 1}$   
and Zhan-Wen Han$^{\rm 1}$\thanks{E-mail:zhanwenhan@ynao.ac.cn}    \\ 
$^1$Yunnan Observatories, Chinese Academy of Sciences, Kunming 650216, China\\
$^2$School of Science, Qingdao University of Technology, Qingdao 266525, China\\
$^3$School of Physics and Electrical Information, Shangqiu Normal University, Shangqiu 476000, China\\
$^4$Physics Department and Tsinghua Center for Astrophysics, Tsinghua University, Beijing 100084, China\\
$^5$School of Astronomy and Space Science, Nanjing University, Nanjing 210023, China}
\begin{document}
\date{Accepted. Received}
\pagerange{\pageref{firstpage}--\pageref{lastpage}} \pubyear{2021}
\maketitle

\label{firstpage}

\begin{abstract}
Ultra-compact X-ray binaries (UCXBs) are low-mass X-ray binaries 
with hydrogen-deficient mass-donors and ultra-short orbital periods.
They have been suggested to be the potential Laser Interferometer Space Antenna (LISA) sources in the low-frequency region.
Several channels for the formation of UCXBs have been proposed so far.
In this article,  we carried out a systematic study on the He star donor channel, in which a 
neutron star (NS) accretes matter from a  He main-sequence (MS) star through Roche-lobe overflow, 
where the mass-transfer is driven by the gravitational wave radiation.
Firstly, we followed the long-term evolution of the NS+He MS  binaries by employing 
the stellar evolution code Modules for Experiments in Stellar
Astrophysics, and thereby obtained the initial parameter spaces for the production of
UCXBs. We then used these results to perform a detailed binary population synthesis 
approach to obtain the Galactic rates of UCXBs through this channel.  
We  estimate the Galactic rates of UCXBs appearing as LISA sources to be $\sim3.1-11.9\, \rm Myr^{-1}$ through this channel,
and the number of such UCXB-LISA sources in the Galaxy can reach about $1-26$ calibrated by observations.
The present work indicates that the He star donor channel may contribute significantly to the Galactic UCXB formation rate.
We found that the evolutionary tracks of UCXBs  through this channel can account for the location of the five transient sources 
with relatively long orbital periods quite well.
We also found that such UCXBs can be identified by their 
locations in the mass-transfer rate versus the orbital period diagram.

\end{abstract}


\section{Introduction} 

Ultra-compact X-ray binaries (UCXBs) are a sub-type of low-mass X-ray binaries (LMXBs),  
containing a compact accretor and a hydrogen-deficient mass-donor with an ultra-short orbital
period (usually less than 1 hour; for a review see Nelemans \& Jonker 2010).  
They are a kind of accretion-powered X-ray sources, in which
the accretor could be a neutron star (NS) or a stellar-mass black hole (BH) that is accreting matter from a mass-donor through
Roche-lobe overflow  (RLOF; e.g. Savonije, de Kool \& van den Heuvel 1986).
So far, there are about 40  known UCXBs or candidates, 
in which 15  of them have been identified with high confidence on the basis of 
the observed orbital periods (including 10 persistent  sources and 5 transient sources; 
see, e.g. in't Zand, Jonker \& Markwardt 2007; Liu, van Paradijs \& van den Heuvel 2007; 
Heinke et al. 2013;  Cartwright et al. 2013; Pietrukowicz et al. 2019; Coti Zelati et al. 2021; Peng \& Shen  2021). 
It has been suggested that  all confirmed UCXBs contain NS accretors,
but the nature of accretors has not been identified exclusively in the observations  (e.g. Sazonov et al. 2020;  
Coti Zelati et al. 2021). \footnote{UCXBs with BH accretors have not been confirmed, 
but there is a known moderately strong BH UCXB candidate 47 Tuc X9 in our Galaxy (see Bahramian et al. 2017; Tudor et al. 2018). }
To date, about a third of observed UCXBs have been found in dense globular clusters that can
enhance UCXB production owing to dynamical interactions, such as  tidal captures or stellar encounters   (see, e.g. Verbunt 1987; 
Bailyn \& Grindlay 1987;  Davies, Benz \& Hills 1992; Davies \& Hansen 1998; Ivanova et al. 2005, 2010).

UCXBs play an important role in broad aspects of astrophysics, as follows: 
(1)  UCXBs  have been thought to be
strong continuous gravitational wave (GW) sources in the low-frequency region 
($\sim10^{-4}-10^{-3}\,\rm Hz$), which can be detected by the  space GW detectors, 
such as Laser Interferometer Space Antenna (LISA; e.g. Nelemans 2009; Amaro-Seoane et al. 2017; Tauris 2018),
Taiji (e.g. Luo et al. 2020; Ruan et al. 2020, 2021), 
and TianQin  (e.g. Luo et al. 2016;  Bao et al. 2019; Wang et al. 2019; Huang et al. 2020), etc. 
(2) UCXBs provide important constraints on  the binary evolution, 
such as the angular-momentum loss mechanisms, the  common-envelope evolution 
and the mass-accretion process of compact objects, etc (e.g. Zhu, L{\"u} \& Wang 2012).
(3) UCXBs are  excellent astrophysical laboratories, since they are interesting X-ray sources  and also the combination of ultra-short orbital
periods, compact accretors and  mass-donors with different chemical compositions  (e.g. Nelemans \& Jonker 2010; Lin \& Yu 2018).
(4) UCXBs have been proposed to be progenitor candidates of millisecond radio pulsars (see Alpar et al. 1982).

Due to the ultra-short orbital periods, the mass-donors in UCXBs could be constrained to be hydrogen-deficient,   
partially or fully degenerate stars, such as white-dwarfs (WDs) or non-degenerate naked He stars  
(e.g. Rappaport, Joss \& Webbink 1982; Nelson, Rappaport \& Joss 1986; Bisnovatyi-Kogan 1989;
Deloye \& Bildsten 2003; Zhong \& Wang 2011).\footnote{Another important ultra-compact systems are AM CVn binaries, 
which have similar mass-donors with those of UCXBs but the mass-accreting objects are carbon-oxygen (CO) or oxygen-neon (ONe) WDs 
(see, e.g. Warner 1995; Podsiadlowski, Han \& Rappaport 2003; Nelemans 2005; Nelemans \& Jonker 2010; Solheim 2010; Liu, Jiang \& Chen 2021).}
The spectra in some UCXBs  indicate that 
the accreted matter onto NSs mainly consists of helium or carbon and oxygen,  
which may help to identify the mass-donors  (e.g. Nelemans et al. 2004; Nelemans, Jonker \& Steeghs 2006). 
Especially, thermonuclear (type I) X-ray bursts on the surface of NSs show distinct features 
in dependence of  the accreted matter,  which can be used to constrain  the compositions of the mass-donors in UCXBs
(see, e.g. Cumming 2003; Galloway et al. 2008).
van Haaften, Voss \& Nelemans (2012) recently obtained the compositions of the mass-donors in 10 UCXBs, 
in which about half of them are likely to be rich in helium.

Up to now, several channels and their variants for the formation of UCXBs in the field have been proposed 
based on the nature of the mass-donors, in which the  mass-donor could be a WD, 
an evolved  main-sequence (MS)  star or a He star  (for a review see Nelemans \& Jonker 2010).\footnote{In globular clusters, 
however, the formation of UCXBs is probably dominated by dynamical interactions (e.g. Verbunt 2005). }
In the classical channel,  a NS accretes matter from a WD through RLOF, 
in which the WD fills its Roche-lobe due to the rapid orbital shrinkage induced 
by GW radiation, known as  the WD donor channel
(see, e.g. Pringle \& Webbink 1975; Tutukov \& Yungelson 1993; Iben, Tutukov \& Yungelson 1995; 
Yungelson, Nelemans \& van den Heuvel 2002; 
Belczynski \& Taam 2004; van Haaften et al. 2012a; Jiang, Chen \& Li 2017; Yu, Lu \& Jeffery 2021). 
Sengar et al. (2017)  performed a number of complete 
binary computations for the formation of UCXBs through stable mass-transfer from He WDs to NSs, 
and suggested that this channel can account for  the properties of the observed UCXBs with high helium abundances.
Bobrick, Davies \& Church (2017) recently argued that only binaries containing He WDs with masses less than  
$0.2\,\rm M_{\odot}$ can undergo stable mass-transfer and then form UCXBs.
By using a newly suggested magnetic braking prescription from Van, Ivanova \& Heinke (2019),
Chen et al. (2021) recently found that the initial orbital period range of LMXBs forming UCXBs 
becomes significantly wider.
It has been suggested that at least one hundred UCXBs will be detected by LISA in the Galaxy 
based on the WD donor channel (see Tauris 2018).

Podsiadlowski, Rappaport \& Pfahl (2002) suggested that 
LMXBs consisting of a NS and an evolved  MS donor with the initial orbital periods 
below the bifurcation period can also evolve 
into UCXBs, known as the MS donor channel.\footnote{LMXBs with the initial orbital periods 
less than the bifurcation period will produce converging systems,  
otherwise they will form diverging systems  (see Pylyser \& Savonije 1988, 1989).}
van der Sluys, Verbunt \& Pols (2005a,b) carried out 
a number of complete binary computations for this channel, 
and explored the bifurcation period of LMXBs that can evolve to UCXBs.
It has been suggested that the bifurcation period is  sensitive to 
the angular-momentum loss mechanisms, 
especially the magnetic braking laws and the mass-loss prescriptions
(see, e.g. Ergma 1996; Podsiadlowski, Rappaport \& Pfahl 2002;  Ma \& Li 2009a; Shao \& Li 2015; 
Chen et al. 2017; Deng et al.  2021).
It is worth noting that  part of the current LMXBs may originate from the evolution of intermediate-mass X-ray binaries  (IMXBs)
that appear as LMXBs in most of their X-ray active lifetime (e.g. Podsiadlowski, Rappaport \& Pfahl 2002; 
Li 2002, 2015; Xu \& Li 2007; Lin et al. 2011; Shao \& Li 2012).
Chen \& Podsiadlowski (2016) argued that the pre-IMXBs with initial orbital periods 
much near the bifurcation period can evolve towards UCXBs via the magnetic braking caused 
by the coupling between the magnetic field and an irradiation-driven wind. 
Chen, Liu \& Wang (2020) recently suggested that 
pre-LMXB/IMXB systems need fine-tuning in the initial orbital period (much near the bifurcation period) to produce detached pre-UCXBs 
(see also Istrate, Tauris \& Langer 2014).
In addition, as a variant of the MS donor channel,
Ma \& Li (2009b) proposed an alternative channel for the formation of UCXBs by considering circumbinary disk-driven mass-transfer.

A NS can also accrete matter from a  non-degenerate He star via RLOF, 
where the mass-transfer is driven by GW radiation, 
known as the He star donor channel (see Savonije, de Kool \& van den Heuvel 1986; Heinke et al. 2013).
Savonije, de Kool \& van den Heuvel (1986) first calculated the evolution of an ultra-compact binary  with a He star donor,
and confirmed that NSs can indeed be accompanied by low-mass He stars.
Heinke et al. (2013)  argued that the He star donor channel may 
account for the formation of three persistent UCXBs with relatively 
long orbital periods (40$-$60\,minutes) and high mass-transfer rates,
but they did not perform  complete binary computations.
By using a binary population synthesis (BPS) approach,  
Zhu, L{\"u} \& Wang (2012) roughly estimated that about 50\%$-$80\% of UCXBs have naked He star donors.
However, the parameter space and the rate for the production of UCXBs through this channel are still highly uncertain.

In Paper I (Chen, Liu \& Wang 2020), 
we carried out a systematic study on the detectability of UCXBs by 
LISA through the MS  donor channel, and roughly estimated that the number of UCXB-LISA sources may reach about 240$-$320 in the Galaxy.
Following the Paper I, the purpose of this article is
to investigate the formation and evolution of UCXBs  through the He star donor channel 
in a systematic way, and then to explore their detectability by LISA, Taiji and TianQin in the Galaxy. 
In Section 2, we describe the numerical code  and methods for binary evolution calculations.
The binary evolutionary results are shown in Section 3. We describe the BPS method and present 
the corresponding results in Section 4. Finally, a discussion is given in Section 5,
and a summary in Section 6.

\section{Binary evolution calculations}

In NS+He star systems with tight orbits, GW radiation makes the orbit shrink until the He stars fill their Roche-lobes.
And then, the He stars start to transfer matter onto the surface of the NSs, 
resulting in the mass increase of NSs. 
To investigate the formation of UCXBs  through the He star donor channel systematically, 
it is necessary to carry out detailed binary evolution computations.
Employing the stellar evolution code Modules for Experiments in Stellar
Astrophysics (MESA, version 12778; see Paxton et al. 2011, 2013, 2015, 2018, 2019), we simulated 
the long-term evolution of NS+He star systems for the formation of UCXBs.
The loss of orbital angular-momentum  from GW
radiation is considered by using a standard formula from Landau \& Lifshitz (1971),
\begin{equation}
{{\rm d}\,\ln J_{\rm GW}\over {\rm d}t} = -{32G^3\over 5c^5}\,{M_{\rm NS} M_2
(M_{\rm NS}+M_2)\over a^4},
\end{equation}
where $M_{\rm NS}$, $M_2$, $a$, $G$ and $c$  are the mass of the NS, 
the mass of the He star donor, the orbital separation of the binary,
the gravitational
constant, and the vacuum speed of light, respectively.

In our simulations, we constructed a grid of binary models for  a typical Population I (Pop I) metallicity $Z=0.02$,
in which the He star models are composed of  $98\%$ helium and $2\%$ metallicity.
For the He star models, we built He stars with zero-age through MESA in idealized methods, in which
we neglect all nuclear reactions during their formation, leading to uniform elemental 
abundance profile from the inside out (see also Wong \& Schwab 2019).
In particular, more realistic stripped He stars (e.g. hot subdwarf O/B/A stars,  horizontal branch stars, etc) 
usually have some amount of 
hydrogen on their surfaces, which makes such stars bigger. The effect of hydrogen in mass transferring 
from He stars was recently discussed in Bauer \& Kupfer (2021).

During the binary evolution, the He star donor transfers some of its matter and angular-momentum to the NS  once it fills its Roche-lobe.
Instead of solving the stellar structure equations of NSs, we set the NSs as point masses
and adopt the prescription of  Tauris \& van den Heuvel (2006) for 
the mass-transfer efficiency with $\alpha=0$, $\beta=0.5$ and $\delta=0$, in which $\alpha$, $\beta$ and $\delta$ are the fractions of
the mass-loss from the vicinity of the He star donor in the form of the stellar wind,   the mass-loss from the vicinity of the NS, and 
the circum-binary co-planar toroid, respectively.\footnote{Similar to previous studies (see Podsiadlowski, Rappaport \& Pfahl 2002; Paper I), 
we arbitrarily set  $\beta$ to be 0.5 in this work. We have computed some models with $\beta=0.7$, 
and found that the $\beta$ value has no significant influence on the  final results.}
Accordingly, we define the mass-accretion rate ($\dot{M}_{\rm acc}$) onto the NS as $\dot{M}_{\rm acc}=(1-\beta)\dot{M}_{\rm 2}$, 
where $\dot{M}_{\rm 2}$ is the mass-transfer rate. If $\dot{M}_{\rm acc}$ is larger than the Eddington accretion rate ($\dot{M}_{\rm Edd}$),
we assume that the unprocessed matter is ejected from the vicinity of the NS, taking away the specific
orbital angular-momentum of the accreting NS (see also Paper I). 
For He accretion, we set $\dot{M}_{\rm Edd}$ to be  $3\times10^{-8}\,\rm M_{\odot}\,yr^{-1}$ (e.g. Dewi et al. 2002; Chen, Li \& Xu 2011).
It has been suggested that the transient behavior in X-ray binaries results from the
thermal-viscous instability of accretion discs (e.g. King, Kolb \& Burderi 1996; Dubus et al. 1999). 
In the present work, we rely on the stability criterion based on the irradiated pure helium discs 
(see, e.g. Lasota, Dubus \& Kruk 2008; Heinke et al. 2013).

If the NS+He star systems evolve into UCXBs with orbital periods $P_{\rm orb}\la1-2$ hours, 
they may potentially be detected by  low-frequency GW detectors like LISA, Taiji and TianQin.
Here, we set the critical orbital period of UCXBs to be 1\,hour (e.g.  Nelemans \& Jonker 2010).
Once the calculated characteristic strain is larger than the LISA sensitivity, 
we assume that the corresponding UCXBs are LISA sources.
In this article,  similar to previous studies (see Tauris 2018; Chen 2020),
we adopted the characteristic strain amplitude of UCXBs based on 4\,yr of LISA observations, written as 
\begin{eqnarray}
h_{\rm c}\approx 2.5\times 10^{-20}\left(\frac{M_{\rm chirp}}{\rm M_{\odot}}\right)^{5/3}\left(\frac{f_{\rm GW}}{\rm mHz}\right)^{7/6}\left(\frac{15~\rm kpc}{d}\right),
\end{eqnarray}
in which $d$ is the distance from the source to the detector, and  the GW frequency is defined as 
 $f_{\rm GW}=2/P_{\rm orb}$ 
 where $P_{\rm orb}$ is the orbital period of the binary. 
 For a simplification, we used a formula related to $f_{\rm GW}$ to describe the 
chirp mass, expressed as  
\begin{equation}
M_{\rm chirp}=\frac{c^{3}}{G}\left(\frac{5\pi^{-8/3}}{96}f_{\rm GW}^{-11/3}\dot{f}_{\rm GW}\right)^{3/5},
\end{equation}
in which $\dot{f}_{\rm GW}$ is the derivative of  GW frequency  (see Tauris 2018). 

In principle, Eq. (3) can only be applied in a detached binary, in which the orbital angular-momentum loss is
fully contributed by GW radiation. However, the mass-transfer between two components of UCXBs and the mass 
loss of the system also have influence on the orbital evolution. Therefore, a dynamical chirp 
mass was defined (see Eq. 4 in Tauris 2018). Especially, the GW frequency will decrease in the orbit expanding stage, resulting
in a negative chirp (e.g. Kremer et al. 2017). 
It is worth noting that  Eq. (3) is still representative of the chirp mass by order of magnitude since the mass-transfer is 
driven by GW radiation and happens on a timescale close to that of GWs.

We incorporated the prescriptions above into MESA and evolved a large number of NS+He MS systems for producing UCXBs, 
thus obtaining a large, dense model grid of binaries. 
In our simulations, the initial parameter ranges of NS+He star systems were chosen, as follows:
(1) The initial masses of the NSs, $M_{\rm NS}^{\rm i}$, are set to be $1.4\,\rm M_{\odot}$. 
(2)  The initial masses of the He star  donors, $M_{\rm
2}^{\rm i}$, range from $0.32\,\rm M_{\odot}$ (the minimum mass for a He star to ignite helium in its centre) 
to $1.20\,\rm M_{\odot}$ (the upper limit on He MS mass for a NS+He MS binary to produce an UCXB in our simulations), where 
we used an equal step of  $\bigtriangleup M_{\rm 2}$ to be $0.1\,\rm M_{\odot}$.
(3) The initial orbital periods
of the binaries, $P_{\rm orb}^{\rm i}$, change from the minimum value (at which a He star with zero-age fills its Roche-lobe)
to the maximum value (at which the He star fills its Roche-lobe when its central He is exhausted), where we set  
$\bigtriangleup \log P_{\rm orb}$ to be 0.1.

\section{Binary evolution results}

We performed  a large number of complete binary evolution computations of 
NS+He star systems for the formation of UCXBs. 
Table 1 lists the main evolutionary properties of some typical NS+He star systems that can evolve into UCXBs.
In this table,  we first mainly explored the effect of different initial He star masses on the final results (see sets $1-5$), and then 
the effect of different initial orbital periods  (see sets $6-11$).

\begin{table*}
\begin{center}
\caption{Selected Evolutionary Properties of  Some Typical UCXBs with Different Initial Secondary Masses and Initial Orbital Periods.
Notes. The columns (from left to right):  the initial mass of the donor star and the initial orbital period; the stellar age at the onset of RLOF;
the minimum orbital period,  the stellar age, the donor mass  and   the GW frequency at the minimum orbital period;
the stellar age,  the donor mass, the carbon mass-fraction at the centre of  the donor star, the NS mass 
and the orbital period when the binary evolution terminates;
the timescale being a persistent UCXB based on an irradiated pure helium disc;
and the timescale that the binary appears to be as a LISA source based on a distance of 15\,kpc.}
\begin{tabular}{ccccccccccccccc}
\hline\hline
Set & $M_2^{\rm i}$ & $\log P_{\rm orb}^{\rm i}$ &  $t_{\rm RLOF}$ & $P_{\rm orb}^{\rm min}$ & $t^{\rm min}$ & $M_2^{\rm min}$ & $f_{\rm LISA}$ & $t^{\rm f}$ & $M_2^{\rm f}$ &  $X_{\rm C}^{\rm f}$ &  $M_{\rm NS}^{\rm f}$ & $P_{\rm orb}^{\rm f}$ & $\bigtriangleup t_{\rm per}$ & $\bigtriangleup t_{\rm LISA}$ \\
 & ($\rm M_{\odot}$)   &  (d)  & (Myr)   & (mins)  & (Myr)    &  ($\rm M_{\odot}$)  &  (mHz) &  (Gyr)  
 & ($\rm M_{\odot}$)  & & ($\rm M_{\odot}$) & (mins)& (Myr)& (Myr)\\
\hline
1&0.32  & $-1.90$ &0.75   & 8.12              & 1.06       &0.293  &4.11                          &6.60        & 0.0054    &    0.008 &1.520    &   73.44  & 46.1 &26.0\\
2&0.40 & $-1.80$ &0.37    & 10.47             &4.15        &0.236   &3.17                       &7.27      & 0.0052   &   0.012 &1.585    &    74.88  & 46.6&30.1\\
3&0.50 & $-1.65$ &1.00   &   10.34             &8.27        &0.247   &3.22                       & 4.72    & 0.0085   &   0.065  &1.636     &     66.24 & 50.0&32.1\\
4&0.60 & $-1.55$ &1.92   &   10.37             &14.22      &0.241   &3.21                       &  3.21    & 0.0101   &  0.170 & 1.678      &   63.78  & 55.7& 37.7\\
5&0.70 & $-1.50$ & 1.95    &  10.34            &  20.45    & 0.232   &3.22                       &  0.02    &  0.1824  &  0.294    &1.645    &  11.10   & $...$ & $...$\\
\hline
6&0.40 & $-1.70$ &      1.71    &       10.24        &    5.31  &   0.242    &  3.25  &           8.91   &  0.0049    &   0.022     &    1.581    &    77.89  &48.1 & 31.1   \\
7&0.40 & $-1.60$ &      4.27   &        10.24        &     7.94    &  0.241   &  3.25 &           3.97    &    0.0091   &     0.040    &    1.579    &    62.19  & 48.7 &  31.4  \\
8&0.40 & $-1.50$ &      8.99   &      10.12          &    12.72       &    0.242   & 3.29  &      4.61  &     0.0087    &   0.070     &   1.578     &    66.82 &   53.2 &  36.0  \\
9&0.40 & $-1.40$ &    17.74   &    10.05      &     21.59      &      0.243  &    3.32     &     3.96 &  0.0094       &    0.124    &    1.577    &   65.39   &  62.6 &  38.3   \\
10&0.40 & $-1.30$ &   33.87  &    10.19    &     37.97      &   0.245     &     3.27     &      2.57     &    0.0112     &    0.199    &     1.578   &    61.23  &  79.8 & 38.6  \\
11&0.40 & $-1.20$ &   63.41  &      10.34          & 68.48 &  0.233    &    3.22  &   0.07   &     0.0953    &   0.346     &    1.541    &   16.47  & $...$  &  $...$  \\
\hline
\end{tabular}
\end{center}
\end{table*}

\subsection{A typical example for binary evolution}

\begin{figure*}
\centerline{\epsfig{file=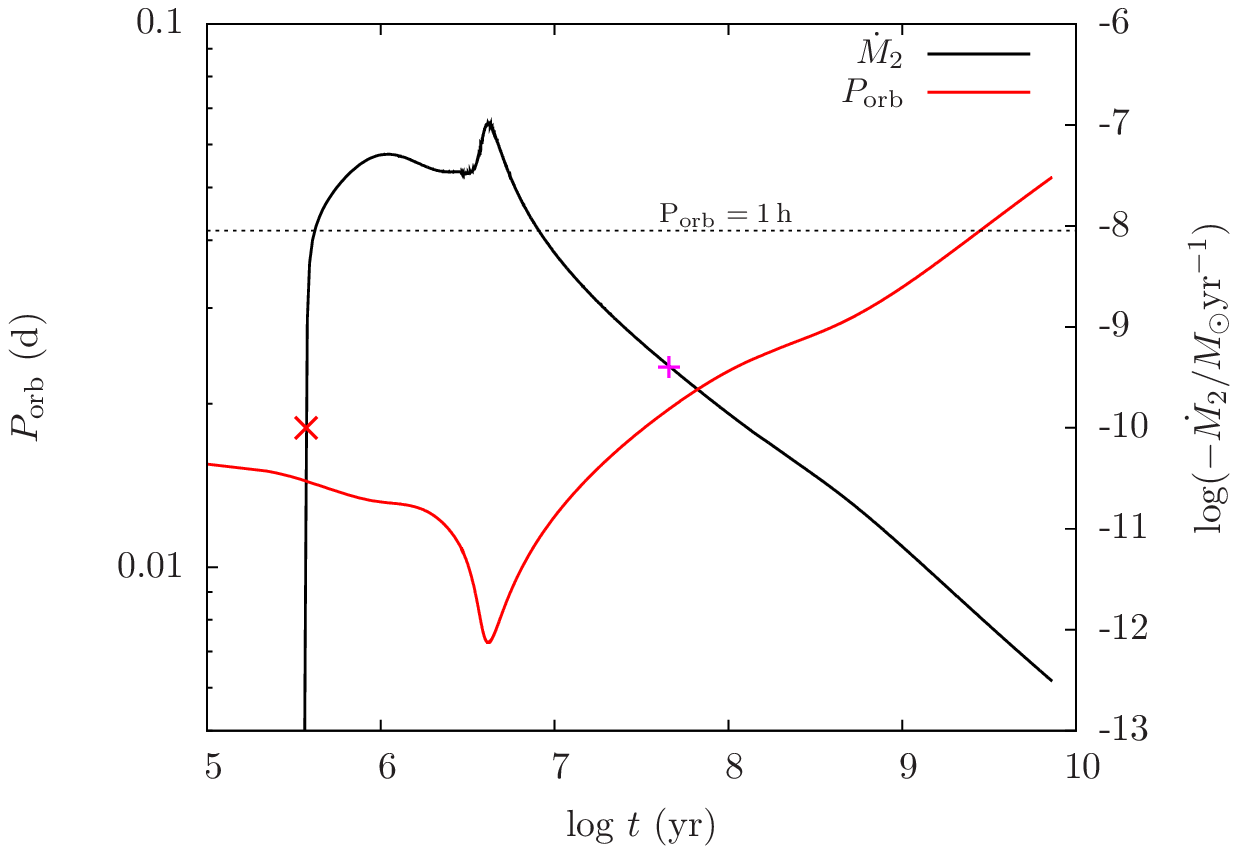,angle=0,width=10cm}\ \
\epsfig{file=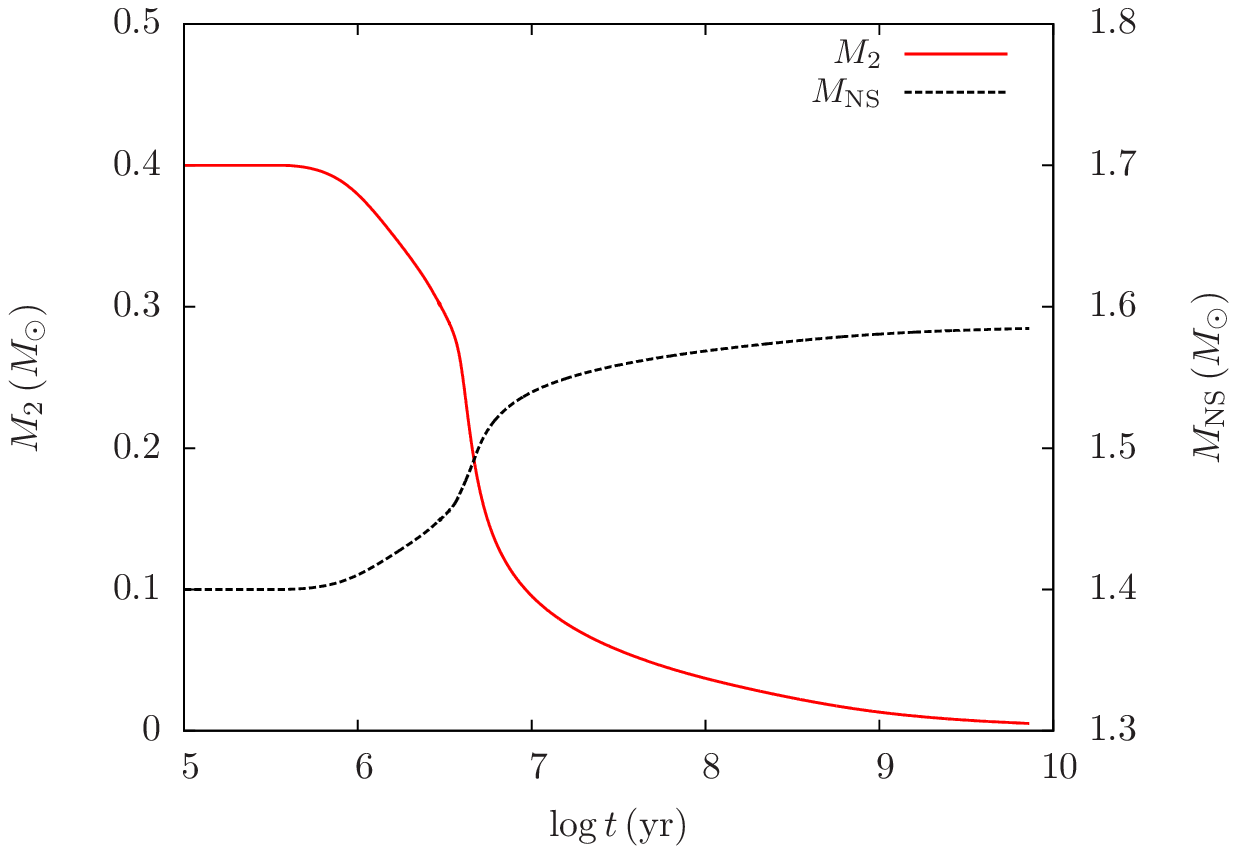,angle=0,width=10cm}} 
\caption{A representative example of the evolution of a NS+He star system that can form an UCXB, 
where ($M_2^{\rm i}$, $M_{\rm NS}^{\rm i}$, $\log (P_{\rm orb}^{\rm i}/{\rm d})$)$=$ (0.40, 1.40, $-1.80$) (see set 2 in Table 1).
Left panel: the evolution of  binary orbital period (red solid curve, left-hand axis)  and
the mass-transfer rate (black solid curve, right-hand axis) as a function of  time for the binary evolution calculations.
The red cross indicates the starting point as a persistent UCXB, whereas the pink plus   represents the end point as a persistent source.
The horizontal black dotted  line shows a critical orbital period of 1\,hour.
Right panel: the evolution of  the mass of the He star donor  (red solid curve, left-hand axis) and the mass of the NS 
(black dashed curve, right-hand axis) as a function of time.}
\end{figure*}

Fig.\,1 shows a typical example of the evolution of a NS+He star system that can form an UCXB (see set 2 in Table 1).
The initial parameters for this binary are
($M_2^{\rm i}$, $M_{\rm NS}^{\rm i}$, $\log (P_{\rm orb}^{\rm i}/{\rm d})$)
$=$ (0.40, 1.40, $-1.80$), in which $M_2^{\rm i}$, $M_{\rm NS}^{\rm i}$ and $P_{\rm orb}^{\rm i}$ are the initial
mass of the He star and the NS in solar mass, and the initial orbital period in days, respectively. 
Due to the short initial orbital period,
the angular-momentum loss induced by the GW radiation is large enough to
result in the rapid shrinking of the orbital separation.
After about 0.37\,Myr, the He star begins to fill its
Roche-lobe while it is still in the He-core burning stage (case BA mass-transfer; see Dewi et al. 2002).
After the beginning of the mass-transfer,  the binary appears to be a persistent UCXB soon.
The binary shows as a persistent source lasting for $\sim$46.6\,Myr.

The binary undergoes severe orbital angular-momentum loss through  
GW radiation and thus keeps much higher mass-transfer rates during the UCXB stage.
After about 4.15\,Myr,  the He star decreases its mass to $0.236\,\rm M_{\odot}$ and
the binary evolves to
the minimum  orbital period  ($P_{\rm orb}^{\rm min}=10.47\,{\rm minutes}$). 
At this stage,  helium burning in the centre of the He star starts to fade.
After that, the binary orbit starts to expand, and the mass-transfer rate begins to decrease
 (for more discussions see Sect. 3.2).  
The binary appears to be an UCXB lasting for  about 2.8\,Gyr based on a critical orbital period of 1\,hour.
When the He star donor decreases its mass to $0.0052\,\rm M_{\odot}$, the evolutionary code 
stops because of hitting the equation of state (EOS) limits ({${\rm max\_} \log Q>5$; a MESA default value), 
where $\log Q =-2\log T + \log\rho+12$, in which $T$ and $\rho$  are the temperature and density in a specific zone of the star, respectively.
At this moment, the lifetime of the binary is 7.27\,Gyr, the mass of the NS is 
$M^{\rm f}_{\rm NS}=1.585\,\rm M_{\odot}$, and the orbital period is 
$P_{\rm orb}^{\rm f}=74.88\,{\rm minutes}$.

\subsection{Evolutionary tracks of UCXBs}

\begin{figure}
\begin{center}
\epsfig{file=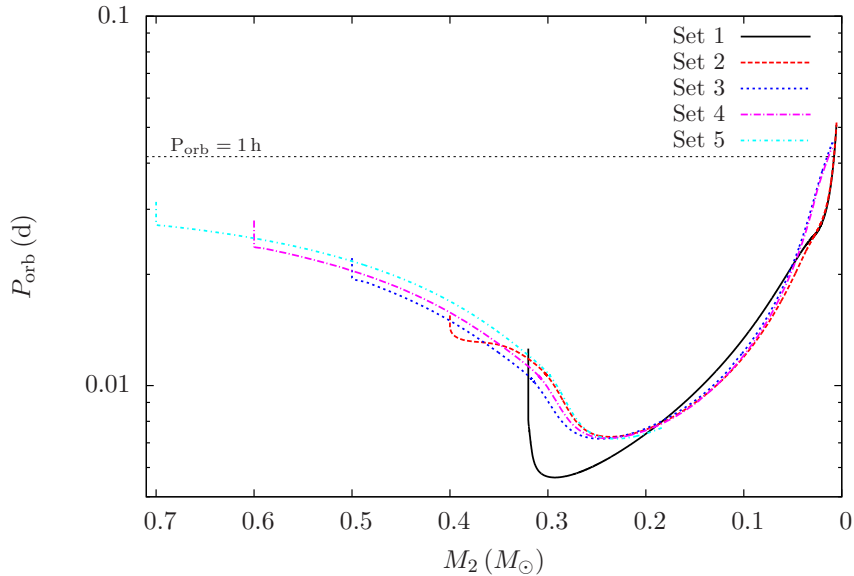,angle=0,width=12cm} 
\caption{Evolutionary tracks of five typical UCXBs with different initial He star masses  (see sets $1-5$ in Table 1)
in the orbital period vs. the secondary mass diagram,  where the horizontal black dotted  line shows a critical orbital period of 1\,hour.}
\end{center}
\end{figure}

Fig. 2 presents the  evolutionary tracks of five typical UCXBs with different initial He star masses  (see sets $1-5$ in Table 1)
in the orbital period versus the secondary mass diagram.
These NS+He star systems would directly evolve into UCXBs once the He star donors fill their Roche-lobes.
During the UCXB stage, the angular-momentum loss is dominated by GW radiation.
When the masses of the He star donors drop below $\sim0.3\,\rm M_{\odot}$, 
the minimum orbital periods of these binaries will be approached soon. The 
minimum orbital periods of UCXBs in the present work are in the range of $8-10$\,minutes, which are 
larger than those  in NS+He WD systems a little bit ($5-7$\,minutes; see, e.g. Sengar et al. 2017; Paper I). 
The main reason is that He WDs are more compact than He star donors. 
At the stage of the minimum orbital periods,  helium burning in the centre of the He star donors is gradually extinguished. 
Meanwhile, 
the He star donors gradually reach a state of mildly degenerate, indicating that their mass-radius exponent is negative.

\begin{figure}
\begin{center}
\epsfig{file=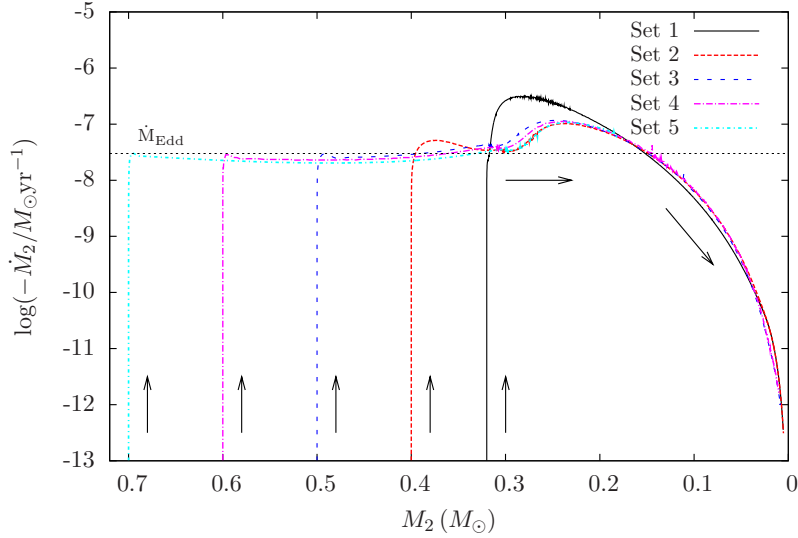,angle=0,width=12cm} 
\caption{Evolutionary tracks of five typical UCXBs with different initial He star masses  (see sets $1-5$ in Table 1)    
in the mass-transfer rate vs. the secondary mass diagram, where 
the tracks follow  the direction of the black arrows. 
The dotted line is $\dot{M}_{\rm Edd}$ for a NS that is accreting He-rich material.}
\end{center}
\end{figure}

In Fig. 2, we note that the NS+He star system with $M_2^{\rm i}=0.32\,\rm M_{\odot}$ has a lower minimum orbital period.
This is because the system contains a slightly more massive donor at the moment of the minimum orbital period (see Table 1), 
which is more compact than that of a less massive one for degenerate stars.
With the evolution of the UCXBs,  the masses of the donors and the mass-transfer rates become 
lower and lower  after the minimum orbital periods (see Fig. 3). 
Meanwhile, the binary orbits gradually widen (see Fig. 2), resulting from 
the mass and radius relation for  degenerate mass-donors whereby they expand in response to mass-loss
(see also Savonije, de Kool \& van den Heuvel 1986; Sengar et al. 2017).
These binaries appear to be UCXBs lasting for  about $1.5-4.0$\,Gyr, where 
they present as persistent sources with lifetime about $40-80$\,Myr.
Note that the NS+He star system represented by set 5 quickly enters into the mass-transfer stage
due to the short orbital period, leading to the formation of an UCXB.  After about 20.45\,Myr, as shown in Fig. 2, the binary
enters into the minimum orbital period, after which we suffer from some numerical difficulties 
for the evolution of the NS system (for a similar case, see  set 11 in Table 1) .

After the UCXB stage, 
the evolved NS systems will form millisecond radio pulsars, in which the pulsars
have been regenerated by mass-accretion induced spin-up. 
If the millisecond pulsars start to evaporate their companions, 
black widows will be formed (see, e.g. van den Heuvel \& van Paradijs 1988; 
Ruderman, Shaham \& Tavani 1989; Chen et al. 2013; Tang \& Li 2021). 
The NS systems studied here will eventually produce single millisecond radio 
pulsars or pulsar+planet-like systems (for more discussions see Sect. 5).

\subsection{The $\dot{M}_{\rm 2}-P_{\rm orb}$ diagram}

\begin{figure}
\begin{center}
\epsfig{file=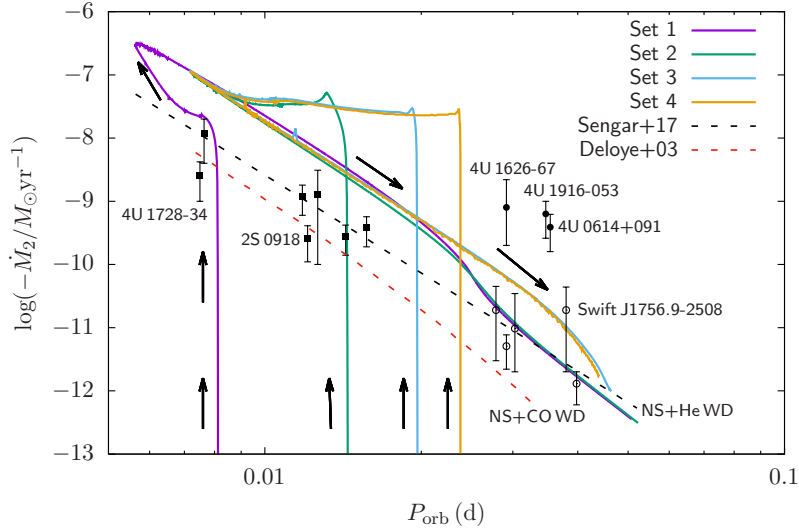,angle=0,width=12cm} 
\caption{Evolutionary tracks of four typical UCXBs with different initial He star masses   (see sets $1-4$ in Table 1)  
in the mass-transfer rate vs. the orbital period diagram, where 
the tracks follow the direction of the black arrows. 
For a comparison, in this figure we also plot the theoretical evolutionary tracks of
the mass-transfer rates on the declining stage based on 
the NS+He WD channel and the NS+CO WD channel.
The black dashed line shows the numerical relation fit between the  mass-transfer rate and the orbital period 
based on the NS+He WD channel (see Sengar et al. 2017),
whereas the red dashed line presents the low-entropy  evolutionary tracks  for degenerate O donors 
based on the NS+CO WD channel  (see Deloye \& Bildsten 2003).
More details of the orbital period and the mass-transfer rate changing with the secondary mass can be found in Figs 2 and 3.
The solid squares, solid circles, and open circles denote seven persistent sources with short orbital periods, 
three persistent sources and  five transient sources with relatively long orbital periods (40$-$60\,minutes), respectively.  
The observational data are taken from Heinke et al. (2013). 
Note that the upper and lower error estimates for the mass-transfer rates are still not well determined,
and the orbital periods of two UCXBs (i.e. 4U 1728-34 and 4U 0614+091) may not be correct in these observational data 
(see Heinke et al. 2013).}
\end{center}
\end{figure}

Fig. 4 represents the evolutionary tracks of four typical UCXBs with different initial He star masses  (see sets $1-4$ in Table 1)     
in the mass-transfer rate versus the orbital period diagram.
The evolutionary tracks in this diagram exist two stages (see also Sengar et al. 2017), as follows:
(1) \textit{The ascending stage}. 
The orbital periods continue to decrease due to GW radiation after the He star donors fill their Roche-lobes, 
whereas the tracks of the mass-transfer rates enter into the ascending stage until the tip  at the minimum orbital periods.
(2) \textit{The declining stage}. 
Following the minimum orbital periods, the tracks of the mass-transfer rates get into the declining stage
while the orbital periods increase. The observed UCXBs are much more likely to be  on the declining stage as  
their evolutionary lifetime are much longer than a Gyr  along this stage (see also Deloye \& Bildsten 2003; 
Nelemans et al. 2010; Sengar et al. 2017).

As shown in Fig. 4,
our UCXB tracks can explain the location of the five transient sources 
with relatively long orbital periods quite well, especially for Swift J1756.9-2508 
that cannot be reproduced by the WD donor  channel.  
For a comparison, in this figure we also plot the numerical relation fit between the  mass-transfer rate and the orbital period 
based on the NS+He WD  channel  (see Sengar et al. 2017) and 
the low-entropy  evolutionary tracks  for degenerate O donors based on the NS+CO WD  channel (see Deloye \& Bildsten 2003).
We note that most of UCXBs are consistent with the evolutionary tracks of the NS+He WD  channel.
Meanwhile, one UCXB (2S 0918-549)  can be reproduced by a low-entropy CO WD donor, 
consistent with  carbon and oxygen lines in its optical spectra (see Deloye \& Bildsten 2003). 

However, three persistent sources that have relatively long orbital periods (40$-$60\,minutes) 
and high mass-transfer rates ($>10^{-10}\,\rm M_{\odot}\,yr^{-1}$), are
inconsistent with  the  tracks of  the WD donor channel (see also Heinke et al. 2013).  
In the three persistent sources, 
one UCXB (4U 1916-053) exhibits strong helium lines in its optical spectra (probably 
containing a He star donor; e.g. Nelemans, Jonker \& Steeghs 2006), and 
two UCXBs (4U 1626-67 and 4U 0614+091) clearly show the lack of hydrogen and helium lines 
but present carbon and oxygen lines in their optical spectra (likely having CO WD donors; 
see, e.g.  Nelemans et al. 2004, 2006; Werner et al. 2006). 
In the present work, the ascending stage
of massive He star donors could pass near 4U 1916-053, but the passage time 
through that region is too short. The descending stage instead 
goes too far below the three persistent sources (see Fig 4). It has been argued that
4U 1626-67 was produced through accretion induced collapse of an ONe WD accretor,  
likely containing a neon-enriched donor  (see Yungelson, Nelemans \& van den Heuvel 2002).
For more discussions on the formation of the three persistent sources, see Sect. 5.

\begin{figure}
\begin{center}
\epsfig{file=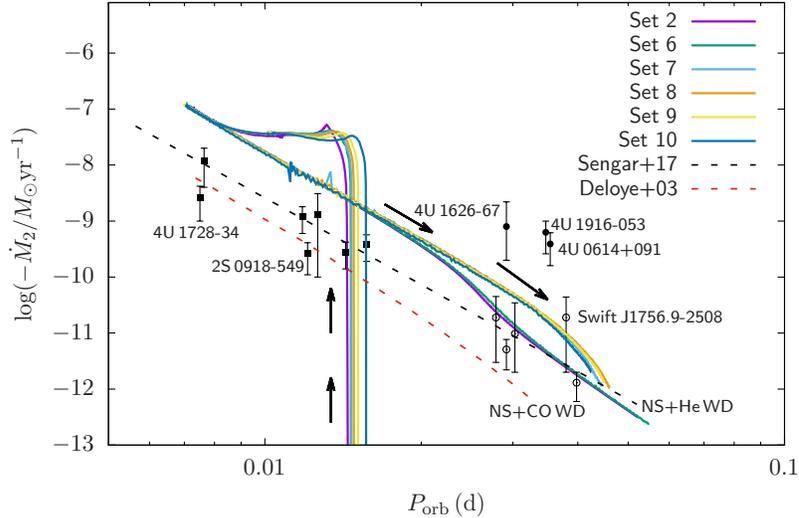,angle=0,width=12cm} 
\caption{Similar to Fig. 4, but for the evolutionary tracks of UCXBs  for $M_2^{\rm i}=0.40\,\rm M_{\odot}$ 
with different initial  orbital periods (see sets 2 and $6-10$ in Table 1).  }
\end{center}
\end{figure}

\begin{figure}
\begin{center}
\epsfig{file=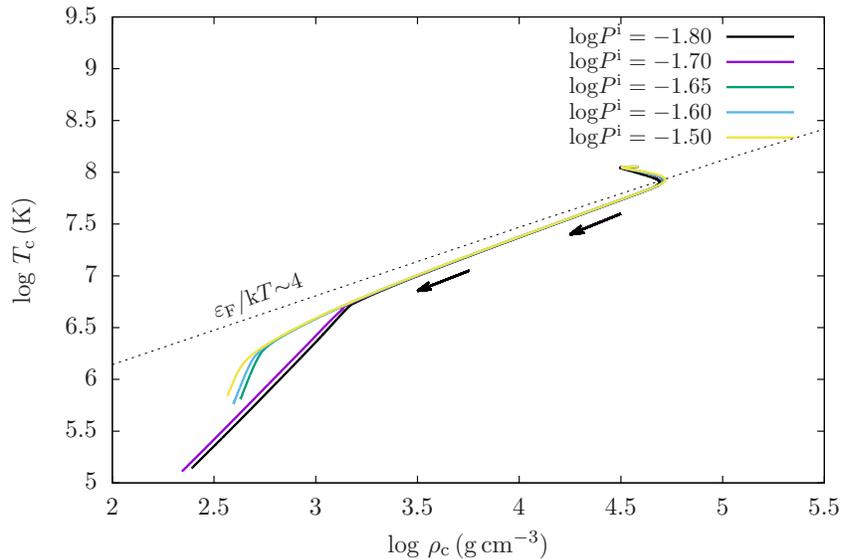,angle=0,width=12cm} 
\caption{Evolutionary tracks of UCXB donor stars for $M_2^{\rm i}=0.40\,\rm M_{\odot}$ with different initial  orbital periods   
in the central temperature vs. the central density diagram, where 
the tracks follow  the direction of the black arrows. 
Non-degenerate and degenerate regions are separated by the dotted line ($\varepsilon_{\rm F}/{\rm   k}{T} \sim 4$).}
\end{center}
\end{figure}

In Fig. 4, we note that the mass-transfer rates on the declining stage have a bifurcation 
after  $P_{\rm orb}>0.02\,{\rm d}$ through the  He star donor channel, where the mass-donors become
lower than $0.06\,\rm M_{\odot}$.
This bifurcation also appears for the cases
with different initial  orbital periods (see Fig. 5).  
Fig. 6 shows the evolutionary tracks of UCXB donor stars for $M_2^{\rm i}=0.40\,\rm M_{\odot}$ with different initial  orbital periods   
in the central temperature versus the central mass density diagram.
As shown in this figure, the bifurcation may be  mainly caused by the central temperature of the mass-donors;
a mass-donor with higher central temperature (higher central carbon mass-fraction) has lower degeneracy, 
resulting in higher mass-transfer rates on the declining stage.

\subsection{Gravitational wave signals}

The observation of GW radiation has started  a new astronomical era after the detection of high-frequency GW signals
from the first double BH coalescence event (GW150914)  by the ground-based aLIGO/Virgo
 (see Abbott et al. 2016).  The second historic moment is
 the GW detection of the first  double NS coalescence event  (GW170817)  together with
the entire electromagnetic spectrum from gamma-rays to radio signals, 
becoming a milestone in multi-messenger astrophysics  (see Abbott et al. 2017). 
UCXBs have been thought to be the potential  GW sources  in the low-frequency region, 
which would be observable  by the future space-based GW observatories 
like LISA, Taiji and TianQin.
Owing to the continuous mass-accretion, 
they provide a favorable advantage to carry out full multi-messenger investigations 
in both electromagnetic spectrum and GW signals.

\begin{figure}
\begin{center}
\epsfig{file=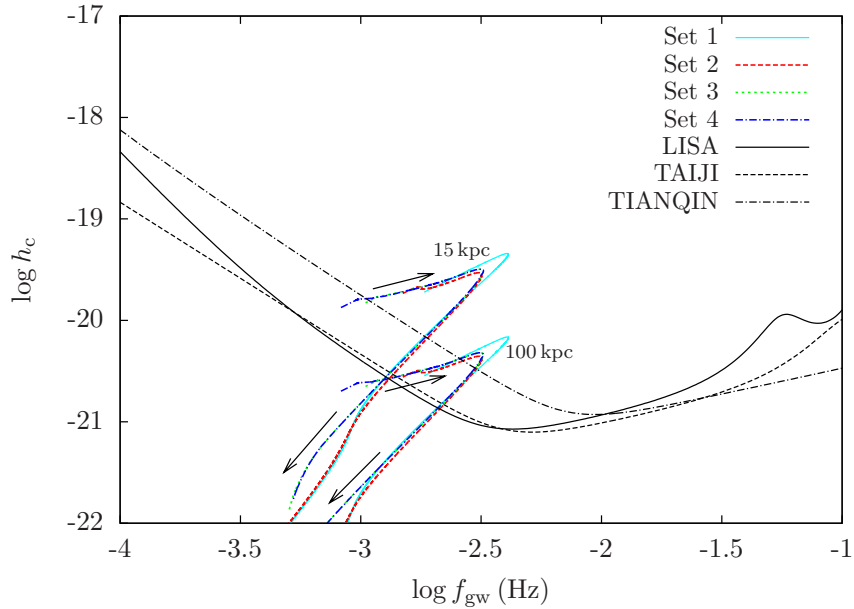,angle=0,width=12cm} 
\caption{Evolutionary tracks of four typical UCXBs with different initial He star masses  (see sets 1$-$4 in Table 1)   
in the characteristic strain amplitude vs. the GW frequency diagram, where 
the tracks follow  the direction of the black arrows. 
The upper and under curves represent the binary evolution results based on the distance of 15\,kpc and 100\,kpc from the source to the detectors, respectively.
The sensitivity curve for the future space-based GW observatory LISA is from the numerical calculations based on 4\,yr of observations (see Robson et al. 2018).
For comparison, in this figure we also present the Taiji sensitivity curve (see Ruan et al. 2020) and the TianQin sensitivity curve (see Wang et al.  2019).  }
\end{center}
\end{figure}

Fig. 7 shows the evolutionary tracks of four typical UCXBs with different initial He star masses  (see sets $1-4$ in Table 1)    
in the characteristic strain amplitude versus the GW frequency diagram.  
We suppose that an UCXB can be visible as a LISA source once  the calculated characteristic strain is larger than the LISA sensitivity.
Note that this assumption only provides an upper limit for the number estimation of UCXB-LISA sources.
From this figure, we can see that
UCXBs through the He star donor channel can be detected by the future space-based GW observatories like LISA, Taiji and TianQin.
Due to the short initial orbital periods, the preceding detached NS+He star binaries will be also detected by the GW observatories and
would be observable as binary radio pulsars.
Meanwhile, the sources will be both seen as persistent UCXBs and GW sources simultaneously after the onset of the mass-transfer.

With the evolution of an UCXB, the peak characteristic strain approaches when the minimum orbital period is coming.
In our simulations,  all NS+He star systems forming  UCXBs
can be detectable by LISA within a distance of 15\,kpc. 
At distances up to the edge of the Galaxy ($\sim$100\,kpc), they are even be visible as LISA sources.
The average $\bigtriangleup t_{\rm LISA}$ within a distance of 15\,kpc is about 33\,Myr in our simulations (see Table 1).
We also note that
most of UCXBs produced by the MS donor channel  can only be visible by LISA within a distance of 1\,kpc (see Paper I).
Compared with the MS donor channel, UCXBs from the He star donor channel are easier to be discovered by LISA.

\subsection{Initial parameters of NS+He star systems}

\begin{figure}
\begin{center}
\epsfig{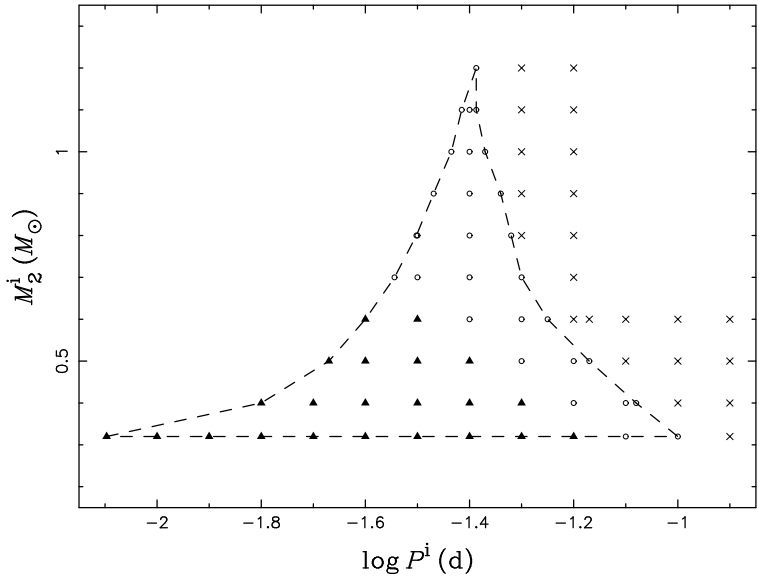} 
\caption{Regions in the initial
orbital period--secondary mass plane ($\log P^{\rm i}$, $M^{\rm
i}_2$) for NS+He star binaries that can produce UCXBs, 
in which the initial mass of NSs is assumed to be $1.4\,\rm M_{\odot}$. 
The filled triangles and open circles stand for the binaries resulting in the formation of UCXBs.
For the binaries represented by the filled triangles, the evolutionary code terminates because of hitting EOS limits.
For the binaries marked by the open circles,  the evolutionary code stops due to the numerical difficulties 
we suffered after the minimum orbital period (e.g. set 5 in Fig. 2).
Crosses are for those that the He star fills its Roche-lobe when it evolves to the subgiant stage.}
\end{center}
\end{figure}

We evolved a large number of NS+He star systems for the formation of UCXBs, 
and thereby obtained a large, dense model grid of binaries. 
Fig. 8 presents the initial contour for producing UCXBs in
the $\log P^{\rm i}-M^{\rm i}_2$ plane,
where $P^{\rm i}$ and $M^{\rm i}_2$ are the initial orbital period and the initial
mass of the He star donor, respectively. If the initial parameters of a
NS+He star system are located in the
contours, an UCXB is then supposed to be formed. Thus, the
contours can be expediently used in BPS simulations to study the formation of UCXBs. 
In our calculations,  the estimated minimum orbital periods  of UCXBs with He star donors 
are close to 8\,minutes (see set 1 in Fig. 2). 
We also found  that NS+He star systems initialised at wider orbits and coming into contact due to GW radiation
would also be observed as UCXBs (see also Fig. 5).
Note that all UCXBs in this contour can be visible as LISA sources 
within a distance of 15\,kpc, and  they will be still detectable 
as LISA sources within a distance of 100\,kpc.

The  initial parameter space for producing UCXBs is constrained by the following conditions:
(1) The left boundary of the contour is set by the
condition that RLOF starts when the He star is on the zero-age MS stage.
(2) The lower boundary is  constrained by the requirements that
the masses of  He stars  in close binaries  should be larger than $0.32\,\rm M_{\odot}$,
below which helium burning in their centre would be extinguished (see also Han et al. 2002; Yungelson 2008).
(3) The systems beyond the right boundary start mass-transfer 
when the He stars are on the subgiant stage and have CO cores.
Note that  systems close to the right boundary can also form UCXBs when the He stars evolve to the subgiant stage,
but we did not consider this case in the present work.

\section{BPS assumptions and results}
\subsection{BPS assumptions}
To investigate the Galactic rates of UCXBs through the He star donor channel, 
a series of Monte Carlo BPS simulations are carried out  based on 
the rapid binary star evolution (BSE) code developed by Hurley, Tout \& Pols (2002).
The principle  assumptions and basic BPS setup in this work are similar to those in Paper I,
but here we adopted the updated initial parameter space of the He star donor channel for producing UCXBs.
In the BPS approach, the primordial binary samples are obtained in the way of
Monte Carlo simulations.  In each simulation, a sample of $1\times10^{\rm 7}$ primordial binaries 
are tracked until the formation of NS+He star systems. 
We assume that an UCXB would be formed when the initial parameters of a NS+He star system are located in the UCXB 
parameter space of  Fig.\,8.  Note that the He stars from the BSE code here are identified with He stars initialised in MESA simulations.

NS+He star systems for producing UCXBs have most likely emerged from the common-envelope (CE)  evolution of giant binaries as they 
have ultra-short orbital periods.
We assume that the mass-transfer would be dynamically unstable and a CE would be formed 
when the mass-ratio of the binary is larger than a critical value ($q_{\rm c}$; see Paczy\'{n}ski 1976). 
It has been generally thought that $q_{\rm c}$  changes with the evolutionary stage of the
primordial primary once it fills its Roche-lobe (e.g. Hjellming \& Webbink 1987; Podsiadlowski, Rappaport \& Pfahl 2002;
Han et al. 2002). 
In this work, we set $q_{\rm c} = 4$ once the primary is in the MS 
or Hertzsprung gap stage (e.g. Han, Tout \& Eggleton 2000; Chen \& Han 2003). 
When the mass donor is a red-giant (RG) star or  an asymptotic giant branch (AGB) star, we adopt
\begin{equation}
q_{\rm c}=[1.67-x+2(\frac{M_{\rm c1}}{M_{\rm 1}})^{\rm 5}]/2.13,
\end{equation}
where $M_{\rm c1}$, $M_{\rm 1}$ and $x={\rm d\,ln}\,R_{\rm 1}/{\rm d\,ln}\,M_{\rm 1}$ 
are the core mass of the donor, the donor mass and the mass-radius exponent of 
the donor, respectively (see Hurley, Tout \& Pols 2002). 

The CE ejection process is still highly uncertain  (e.g. Ivanova et al. 2013; Soker 2013; Soker, Grichener \& Sabach 2018; Kruckow et al. 2021).
As in our previous studies (see Wang et al. 2009), the standard energy prescription is 
adopted to calculate the CE evolution (see Webbink 1984),  
in which we combine the CE ejection efficiency ($\alpha_{\rm CE}$) and the stellar structure parameter ($\lambda$)  
into a free parameter $\alpha_{\rm CE}\lambda$ and set it to  be 1.0 as our standard model.
Note that other energy sources (e.g. the internal
energy of the envelope) may also  contribute to the CE ejection  (see Han, Podsiadlowski \& Eggleton 1995).
Thus, we consider the effect of a larger value of  $\alpha_{\rm CE}\lambda$  (i.e. 1.5) on the final results for a comparison.

The following input is adopted  in our BPS Monte Carlo simulations
(for a recent review see Han et al. 2020; see also Han 2008; Wang, Podsiadlowski \& Han  2017):
(1) We used the initial mass function  from Miller \& Scalo (1979). 
(2) All stars are assumed to be members of binaries, and 
a constant mass-ratio distribution is adopted. 
(3) A circular orbit is  supposed for the primordial binaries.
(4) The orbital separation distribution is supposed to be constant in $\log a$ for wide primordial binaries,
in which $a$ is the orbital separation of the binary. 
(5) A constant star formation rate (SFR) is assumed over the past 15\,Gyr, 
in which we  suppose that a primordial binary 
with its primary $>0.8\,\rm M_{\odot}$ is produced every year  (see
Han, Podsiadlowski \& Eggleton 1995; Hurley, Tout \& Pols 2002).  
From this calibration, a constant SFR of $5\,\rm M_{\odot}{yr}^{-1}$ can be obtained (see Willems \& Kolb 2004).

\subsection{BPS results}

\begin{figure}
\begin{center}
\epsfig{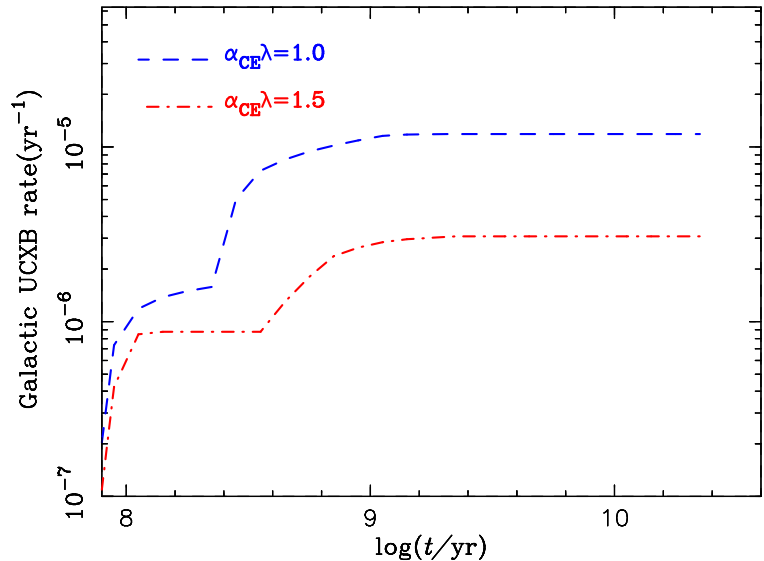} 
\caption{Evolution of the Galactic rates  of UCXBs from the He star donor channel as  a function of time 
by adopting  a constant Pop I SFR of $5~\rm M_{\odot} yr^{\rm -1}$.}
\end{center}
\end{figure}

Fig. 9 shows the evolution of the Galactic rates  of  UCXBs for the He star donor channel
by adopting a constant Pop I  SFR of  $5\,\rm M_{\odot} yr^{\rm -1}$.
The simulations give the  theoretical rates of He star UCXBs  in the Galaxy  to be $\sim3.1-11.9\, \rm Myr^{-1}$.
If we adopt the timescale of UCXBs appearing as LISA sources to be 33\,Myr within a distance of 15\,kpc  (see Table 1), 
there exist about $100-390$ UCXB-LISA sources in the Galaxy, 
at least providing an upper limit for their number estimation. 
The present work indicates that 
the He star donor channel has significant contribution to the formation of  UCXBs.

It is worth noting that there are about 6 persistent UCXBs in the field (e.g. Heinke et al. 2013). 
Persistent UCXBs are believed to be detectable in the whole Galaxy, which
have lifetimes of about $30-100$\,Myr (e.g. Tauris 2018; paper I). Thus, the empirical observed formation rates 
for persistent UCXBs are about $6/(30-100) \,\rm Myr^{-1}=(0.06-0.2)\, \rm Myr^{-1}$.
The theoretical UCXB rates for the He star donor channel are about $15-200$ times larger than the observational values.
Accordingly, calibrating by the UCXB observations, we roughly estimate that
the expected number for such UCXB-LISA sources in the Galaxy is about  $1-26$  (calculated here as $100/200-390/15$).

It seems that the theoretical UCXB rates from the He star donor channel  are far more above the observed rates. 
The main reason for such a discrepancy is due to some uncertainties in our BPS simulations, as follows:
(1) Other Galactic models have already predicted several times smaller SFRs. For example,
Licquia \& Newman (2014) recently suggested a lower Galactic SFR of $1.66\,\rm M_{\odot} yr^{-1}$
based on a hierarchical Bayesian statistical analysis. If this SFR is adopted, the theoretical UCXB rates will
decrease to be $\sim1.0-4.0\times10^{-6} \rm yr^{-1}$ in the Galaxy.
(2)  Recent simulations imply that the critical mass-ratio (i.e. the stability criteria) for RG/AGB mass-transfer is likely to be larger than the adopted value here 
 (e.g. Ge et al. 2020). This indicates that we may slightly overestimate the rates of NS$+$He star systems.

As shown in Fig. 9,  there exist two obvious ascending stages for the evolution of UCXB rates, 
i.e. with the delay time around 100\,{\rm Myr} and 300\,{\rm Myr}
from the star formation to the formation of UCXBs.
These two cases correspond to two specific evolutionary scenarios for the formation of NS+He star systems, in which
the primordial primaries (i.e. the progenitors of NSs) first fill their Roche-lobes and then form CEs
 (1) at the RG stage for the case with short delay times
and (2)  at the AGB stage for the case with long delay times.
For scenario (1), the initial parameters of the primordial binaries are in the range of $M_{\rm 1,i}\sim10-25\,\rm  M_{\odot}$, 
$q=M_{\rm2,i}/M_{\rm1,i}\sim0.1-0.3$ and $P^{\rm i}\sim20-2000\,\rm days$, 
in which $M_{\rm 1,i}$, $M_{\rm 2,i}$, $P^{\rm i}$ and $q$ are the initial masses
of the primordial primary and secondary, the initial orbital period and mass-ratio 
of the primordial systems, respectively. About 15\% of UCXBs from the He star donor channel are produced through this scenario. 
For scenario (2), the initial parameters of the primordial binaries are in the range of 
$M_{\rm 1,i}\sim7-15\,\rm M_{\odot}$, $q\sim0.2-0.8$ and 
$P^{\rm i}\sim600-4000\,\rm days$. About 85\% of UCXBs 
from the He star donor channel are formed through this scenario.

\section{Discussion}

It has been argued that the He star donor channel may produce  
higher average mass-transfer rates in the three persistent UCXBs with relatively 
long orbital periods if the He star donors can retain their initial high entropy (see Heinke et al. 2013).
In the present work,  we found that 
a He star donor with higher central temperature (higher central carbon mass-fraction) has lower degeneracy, 
corresponding to higher mass-transfer rates on the declining stage. Note that we did not consider the irradiation 
effect of X-ray luminosities of mass-accreting NSs on the evolution of UCXBs, which 
may also trigger higher mass-transfer rates (see, e.g. van Haaften, Voss \& Nelemans 2012; Jia \& Li 2015; L\"{u} et al. 2017).
Thus, we speculate that the He star donor channel has the potential ability to account for the formation of the three persistent 
sources with higher mass-transfer rates, especially for 4U 1916-053 that probably has a He star donor.
If the three persistent sources really originate from the He star donor channel, 
we expect  the initial He star mass larger than $0.6\,\rm M_{\odot}$.
In this work, we suffer from some numerical difficulties  for the evolution of NS systems 
with the central carbon abundances of the mass-donors higher than 20\% when the binary orbits gradually widen (see, e.g. set 5 in Fig. 2).
Complete binary computations on these systems are encouraged to unveil the origin of 
the three persistent sources.

As shown in Fig. 4,
different  channels for the formation of UCXBs predict distinct mass-transfer rates on the declining stage
owing to different degeneracy for mass-donors;
the mass-donors with lower degeneracy correspond to higher mass-transfer rates on the declining stage.
Thus, we emphasize  that 
the  relation in the diagram between the mass-transfer rate on the declining stage and the orbital period  
can be used to distinguish the mass-donors in UCXBs.
Meanwhile, different UCXB formation channels have different mass-donor chemical compositions, which 
can be used to provide support for the mass-donor identifications (e.g. Koliopanos et al. 2021). Hence,
more observations on the chemical compositions of the mass-donors in UCXBs are needed.

For the three formation channels of UCXBs, the total UCXB lifetime is mainly 
determined by the initial orbital periods of pre-UCXB systems. 
If we set the critical orbital period of UCXBs to be 1\,hour, 
the total UCXB lifetime is about $2-4$\,Gyr; we found that 
the total UCXB lifetime is no big difference for the three channels (see also Tauris 2018; paper I).
In addition, persistent UCXBs from the He star donor channel have lifetimes of $40-80$\,Myr, 
which are shorter than those from the MS donor channel ($50-150$\,Myr; see paper I), 
but longer than those from  the He WD  donor channel  ($20-40$\,Myr; e.g. Tauris 2018). 
Furthermore,  compared with the WD/MS donor channels,  the He star donor channel has higher possibility to form 
millisecond radio pulsars as  their typical delay times from the star formation to the formation of UCXBs
are shorter than those from the WD/MS donor channels. 

In the present work, the He star donors can decrease their masses to $\sim0.005\,\rm M_{\odot}$
when the binary evolution terminates because of hitting EOS limits.
Ruderman \& Shaham (1985) suggested that NSs  will tidally disrupt the very 
low-mass He degenerate donors before they decrease to
$0.004\,\rm M_{\odot}$  as the mass-donors change their mass-radius relation,  
eventually forming single millisecond pulsars. 
We also note that the NS+He star systems  have the potential possibilities  
to leave behind pulsar+planet-like systems at the end of their evolution (e.g. Podsiadlowski 1993).
Pulsar timing observations have revealed  three pulsar+planet-like systems so far, in which
all planet-like objects are around old millisecond pulsars  (e.g. Martin, Livio \& Palaniswamy 2016).
Especially,  Bailes et al. (2011)  reported a planet-like object around the millisecond pulsar PSR J1719-1438, and 
suggested that this system may have once been an UCXB, in which the mass-donor narrowly 
avoided complete destruction (see also van Haaften et al. 2012b).

UCXBs are important continuous GW sources in the low-frequency region. 
Some recent studies carried out systematic studies on the detectability of UCXBs by 
LISA through the WD/MS donor channels, and suggested that 
several hundred UCXB-LISA sources would be detected  in the Galaxy (see Tauris 2018; paper I).
In the present work, 
the preceding detached NS+He star systems can be detected by the GW observatories owing to 
the short initial orbital periods (see Fig. 7), and they would be also
observable as binary radio pulsars before the mass-transfer process.\footnote{Actually, 
NSs with an ellipticity can also possibly appear as high-frequency GW sources, i.e. 
the detached NS+He star systems have an opportunity to become
dual-line GW sources (see Tauris 2018; Chen 2021; Suvorov 2021).} 
Meanwhile, the preceding detached NS+He star systems could be detectable 
optically as He stars are relatively luminous objects.
By considering the contribution of the WD/MS donor channels (e.g. Tauris 2018; paper I), 
the He star donor channel approximately accounts for one third of UCXBs theoretically on the basis of the present work.
This indicates that we cannot ignore the contribution of the He star donor channel in forming UCXBs.
We note that accretion-induced collapse  of ONe WDs can also produce NS systems, 
likely contributing to the formation of UCXBs (for a recent review 
see Wang \& Liu 2020; see also Tauris et al. 2013; Ablimit \& Li 2015; Wang  2018; Ablimit 2019).

Aside from the contribution to the formation of UCXBs, NS+He star systems also produce 
intermediate-mass binary pulsars (IMBPs). Chen \& Liu (2013) studied NS+He star systems to 
form IMBPs with short orbital periods ($<3$\,d),
in which a He star transfers its material onto the surface of a NS when it
evolves to the subgiant stage, resulting in a recycling process for the NS. Tang, 
Liu \& Wang (2019) recently found that NS+He star systems can reproduce the 
observed parameters of PSR\,J0621$+$1002 that is one of the well-observed IMBPs.
It is worth noting  that
NS+He star systems close to the right boundary in Fig. 8 can also form UCXBs when the He stars evolve to the subgiant stage,
turning into  IMBPs (detached NS+CO WD binaries) at the end. 
The detached NS+CO WD binaries then may spiral in,  experience unstable mass-transfer, and merge, 
producing transient supernova-like events (e.g. Bobrick, Davies \& Church 2017).
If the mass-transfer is stable from CO WDs to NSs,  the binaries will experience UCXB phase again.

In addition, Shao, Li \& Dai (2019)  suggested that NS+He star systems can  significantly 
contribute to the population of ultraluminous X-ray sources  if the NS accretes 
He-rich material with a super-$\dot{M}_{\rm Edd}$ (see also Abdusalam et al. 2020).
Furthermore,  Ma et al. (2020) recently studied type I X-ray bursts
from NSs that accrete He-rich matter, resulting in intermediate X-ray bursts in observations.
Compared with normal X-ray bursts, Ma et al. (2020) found that the intermediate X-ray bursts 
have longer recurrence timescales and higher luminosities.

\section{Summary}

The origin of UCXBs is still highly uncertain.  
In the present work, by combining detailed stellar evolution calculations and BPS approach,
we simulated the formation and evolution of UCXBs  through the He star donor channel in a systematic way.
We firstly explored the parameter spaces for producing UCXBs in the orbital period--secondary mass plane
using detailed binary evolution calculations, and then
used these results to carry out a  series of Monte Carlo BPS simulations. 
The main conclusions are summarized as follows:
\begin{enumerate}
\item [(1)] 
We found that all NS+He star systems for producing UCXBs can be visible as the LISA sources within a distance of 15\,kpc. We
estimate the rates of UCXB-LISA sources in the Galaxy  $\sim3.1-11.9\, \rm Myr^{-1}$. Calibrating by the UCXB observations,
we expect that the number  of UCXB-LISA sources can reach about $1-26$ in the Galaxy. 
\item [(2)] 
The evolutionary tracks of UCXBs  through the He star donor channel are able to explain the location of the five transient sources 
with relatively long orbital periods quite well, especially for Swift J1756.9-2508.  
\item [(3)] 
We found that the relation in the diagram between  the mass-transfer rate on the descending stage and the orbital period  
can be used to distinguish different channels for the formation of UCXBs.
\item [(4)] 
The preceding detached NS+He star systems forming UCXBs, can be detected by the future GW observatories like LISA, Taiji and TianQin.
They would be also observable as binary radio pulsars.
\item [(5)] 
When the He stars fill their Roche-lobes,  the NS+He star binaries appear to be UCXBs and GW sources simultaneously.
The NS+He star systems appear to be UCXBs with lifetime  about $1.5-4.0$\,Gyr, in which they show as 
persistent sources lasting for about $40-80$\,Myr. 
\item [(6)] 
After the UCXB stage, the evolved NS systems studied in this work will show as millisecond radio pulsars, 
eventually forming single millisecond radio pulsars or pulsar+planet-like systems.
\end{enumerate}

\section*{Acknowledgments}
We acknowledge the anonymous referee for valuable comments 
that help to improve the paper.
This study is partly supported by the NSFC (Nos 11521303, 12090040/3, 11573016, 11733009, 
11873085, 12073071, 11903075 and 12003013), 
the science research grants from the China Manned Space Project (Nos CMS-CSST-2021-A13/B07), 
the Program for Innovative Research Team (in Science and Technology) at the
University of Henan Province,
the Youth Innovation Promotion Association of CAS (Nos 2018076 and 2021058),
the Western Light Youth Project of CAS,
and the Yunnan Fundamental Research Projects (Nos 202001AT070058, 2019FJ001 and 202001AS070029).

\section*{Data availability}
The data of the numerical calculations in this  article can be made available on request by contacting BW.

\label{lastpage}
\end{document}